\documentclass[aps,prd,reprint,superscriptaddress,longbibliography,showpacs,showkeys]{revtex4-1}

\usepackage{graphicx}
\usepackage{lineno}
\usepackage{amsmath}
\usepackage{upgreek}
\usepackage[dvips]{color}
\modulolinenumbers[5]
\linenumbers

\begin{document}

\title{Ankle-like Feature in the Energy Spectrum of Light Elements of Cosmic Rays Observed with KASCADE-Grande 
}

\author{W.D.~Apel}
\affiliation{Institut f\"ur Kernphysik, KIT - Karlsruher Institut f\"ur Technologie, Germany}
\author{J.C.~Arteaga-Vel\`azquez}
\affiliation{Universidad Michoacana, Instituto de F\'{\i}sica y Matem\'aticas, Morelia, Mexico}
\author{K.~Bekk}
\affiliation{Institut f\"ur Kernphysik, KIT - Karlsruher Institut f\"ur Technologie, Germany}
\author{M.~Bertaina}
\affiliation{Dipartimento di Fisica, Universit\`a degli Studi di Torino, Italy}
\author{J.~Bl\"umer}
\affiliation{Institut f\"ur Kernphysik, KIT - Karlsruher Institut f\"ur Technologie, Germany}
\affiliation{Institut f\"ur Experimentelle Kernphysik, KIT - Karlsruher Institut f\"ur Technologie, Germany}
\author{H.~Bozdog}
\affiliation{Institut f\"ur Kernphysik, KIT - Karlsruher Institut f\"ur Technologie, Germany}
\author{I.M.~Brancus}
\affiliation{National Institute of Physics and Nuclear Engineering, Bucharest, Romania}
\author{E.~Cantoni}
\altaffiliation{now at: Istituto Nazionale di Ricerca Metrologia, INRIM, Torino}
\affiliation{Dipartimento di Fisica, Universit\`a degli Studi di Torino, Italy}
\affiliation{Osservatorio Astrofisico di Torino, INAF Torino, Italy}
\author{A.~Chiavassa}
\affiliation{Dipartimento di Fisica, Universit\`a degli Studi di Torino, Italy}
\author{F.~Cossavella}
\altaffiliation{now at: Max-Planck-Institut Physik, M\"unchen, Germany}
\affiliation{Institut f\"ur Experimentelle Kernphysik, KIT - Karlsruher Institut f\"ur Technologie, Germany}
\author{K.~Daumiller}
\affiliation{Institut f\"ur Kernphysik, KIT - Karlsruher Institut f\"ur Technologie, Germany}
\author{V.~de~Souza}
\affiliation{Universidade S$\tilde{a}$o Paulo, Instituto de F\'{\i}sica de S\~ao Carlos, Brasil}
\author{F.~Di~Pierro}
\affiliation{Dipartimento di Fisica, Universit\`a degli Studi di Torino, Italy}
\author{P.~Doll}
\affiliation{Institut f\"ur Kernphysik, KIT - Karlsruher Institut f\"ur Technologie, Germany}
\author{R.~Engel}
\affiliation{Institut f\"ur Kernphysik, KIT - Karlsruher Institut f\"ur Technologie, Germany}
\author{J.~Engler}
\affiliation{Institut f\"ur Kernphysik, KIT - Karlsruher Institut f\"ur Technologie, Germany}
\author{M.~Finger}
\affiliation{Institut f\"ur Experimentelle Kernphysik, KIT - Karlsruher Institut f\"ur Technologie, Germany}
\author{B.~Fuchs}
\affiliation{Institut f\"ur Experimentelle Kernphysik, KIT - Karlsruher Institut f\"ur Technologie, Germany}
\author{D.~Fuhrmann}
\altaffiliation{now at: University of Duisburg-Essen, Duisburg, Germany}
\affiliation{Fachbereich Physik, Universit\"at Wuppertal, Germany}
\author{H.J.~Gils}
\affiliation{Institut f\"ur Kernphysik, KIT - Karlsruher Institut f\"ur Technologie, Germany}
\author{R.~Glasstetter}
\affiliation{Fachbereich Physik, Universit\"at Wuppertal, Germany}
\author{C.~Grupen}
\affiliation{Department of Physics, Siegen University, Germany}
\author{A.~Haungs}
\affiliation{Institut f\"ur Kernphysik, KIT - Karlsruher Institut f\"ur Technologie, Germany}
\author{D.~Heck}
\affiliation{Institut f\"ur Kernphysik, KIT - Karlsruher Institut f\"ur Technologie, Germany}
\author{J.R.~H\"orandel}
\affiliation{Dept. of Astrophysics, Radboud University Nijmegen, The Netherlands}
\author{D.~Huber}
\affiliation{Institut f\"ur Experimentelle Kernphysik, KIT - Karlsruher Institut f\"ur Technologie, Germany}
\author{T.~Huege}
\affiliation{Institut f\"ur Kernphysik, KIT - Karlsruher Institut f\"ur Technologie, Germany}
\author{K.-H.~Kampert}
\affiliation{Fachbereich Physik, Universit\"at Wuppertal, Germany}
\author{D.~Kang}
\affiliation{Institut f\"ur Experimentelle Kernphysik, KIT - Karlsruher Institut f\"ur Technologie, Germany}
\author{H.O.~Klages}
\affiliation{Institut f\"ur Kernphysik, KIT - Karlsruher Institut f\"ur Technologie, Germany}
\author{K.~Link}
\affiliation{Institut f\"ur Experimentelle Kernphysik, KIT - Karlsruher Institut f\"ur Technologie, Germany}
\author{P.~{\L}uczak}
\affiliation{National Centre for Nuclear Research, Department of Cosmic Ray Physics, Lodz, Poland}
\author{M.~Ludwig}
\affiliation{Institut f\"ur Experimentelle Kernphysik, KIT - Karlsruher Institut f\"ur Technologie, Germany}
\author{H.J.~Mathes}
\affiliation{Institut f\"ur Kernphysik, KIT - Karlsruher Institut f\"ur Technologie, Germany}
\author{H.J.~Mayer}
\affiliation{Institut f\"ur Kernphysik, KIT - Karlsruher Institut f\"ur Technologie, Germany}
\author{M.~Melissas}
\affiliation{Institut f\"ur Experimentelle Kernphysik, KIT - Karlsruher Institut f\"ur Technologie, Germany}
\author{J.~Milke}
\affiliation{Institut f\"ur Kernphysik, KIT - Karlsruher Institut f\"ur Technologie, Germany}
\author{B.~Mitrica}
\affiliation{National Institute of Physics and Nuclear Engineering, Bucharest, Romania}
\author{C.~Morello}
\affiliation{Osservatorio Astrofisico di Torino, INAF Torino, Italy}
\author{J.~Oehlschl\"ager}
\affiliation{Institut f\"ur Kernphysik, KIT - Karlsruher Institut f\"ur Technologie, Germany}
\author{S.~Ostapchenko}
\altaffiliation{now at: University of Trondheim, Norway}
\affiliation{Institut f\"ur Kernphysik, KIT - Karlsruher Institut f\"ur Technologie, Germany}
\author{N.~Palmieri}
\affiliation{Institut f\"ur Experimentelle Kernphysik, KIT - Karlsruher Institut f\"ur Technologie, Germany}
\author{M.~Petcu}
\affiliation{National Institute of Physics and Nuclear Engineering, Bucharest, Romania}
\author{T.~Pierog}
\affiliation{Institut f\"ur Kernphysik, KIT - Karlsruher Institut f\"ur Technologie, Germany}
\author{H.~Rebel}
\affiliation{Institut f\"ur Kernphysik, KIT - Karlsruher Institut f\"ur Technologie, Germany}
\author{M.~Roth}
\affiliation{Institut f\"ur Kernphysik, KIT - Karlsruher Institut f\"ur Technologie, Germany}
\author{H.~Schieler}
\affiliation{Institut f\"ur Kernphysik, KIT - Karlsruher Institut f\"ur Technologie, Germany}
\author{S.~Schoo}
\altaffiliation{corresponding author: sven.schoo@kit.edu}
\affiliation{Institut f\"ur Experimentelle Kernphysik, KIT - Karlsruher Institut f\"ur Technologie, Germany}
\author{F.G.~Schr\"oder}
\affiliation{Institut f\"ur Kernphysik, KIT - Karlsruher Institut f\"ur Technologie, Germany}
\author{O.~Sima}
\affiliation{Department of Physics, University of Bucharest, Bucharest, Romania}
\author{G.~Toma}
\affiliation{National Institute of Physics and Nuclear Engineering, Bucharest, Romania}
\author{G.C.~Trinchero}
\affiliation{Osservatorio Astrofisico di Torino, INAF Torino, Italy}
\author{H.~Ulrich}
\affiliation{Institut f\"ur Kernphysik, KIT - Karlsruher Institut f\"ur Technologie, Germany}
\author{A.~Weindl}
\affiliation{Institut f\"ur Kernphysik, KIT - Karlsruher Institut f\"ur Technologie, Germany}
\author{J.~Wochele}
\affiliation{Institut f\"ur Kernphysik, KIT - Karlsruher Institut f\"ur Technologie, Germany}
\author{M.~Wommer}
\affiliation{Institut f\"ur Kernphysik, KIT - Karlsruher Institut f\"ur Technologie, Germany}
\author{J.~Zabierowski}
\affiliation{National Centre for Nuclear Research, Department of Cosmic Ray Physics, Lodz, Poland}

\collaboration{KASCADE-Grande Collaboration}
\homepage{www-ik.fzk.de/KASCADE\_home.html}
\noaffiliation

\date{\today}

\begin{abstract}
Recent results of the KASCADE-Grande experiment provided evidence for a mild knee-like structure in the
all-particle spectrum of cosmic rays at $E = 10^{16.92 \pm 0.10} \, \mathrm{eV}$, which was found to be due to a steepening in the 
flux of heavy primary particles. The spectrum of the combined components of light and intermediate masses was found to be compatible with a single power law in the energy range from $10^{16.3} \, \mathrm{eV}$ to $10^{18} \, \mathrm{eV}$. 
In this paper, we present an update of this analysis by using data with increased statistics, originating both from a larger data set including more recent measurements and by using a larger fiducial area. In addition, optimized selection criteria for enhancing light primaries are applied. We find a spectral feature for light elements, namely a hardening at $E = 10^{17.08 \pm 0.08} \, \mathrm{eV}$ with a change of the power law index from $-3.25 \pm 0.05$ to $-2.79 \pm 0.08$. 
\end{abstract}


\maketitle

High energy cosmic rays exhibit a power law like energy spectrum with some features, which have been subject to investigations for many years. For energies above $10^{14} \, \mathrm{eV}$ the flux gets too low to be measured directly, therefore one has to extract information about the primary particles from extensive air showers (EAS), for example, by measuring the secondary particles with ground-based detector arrays. The most prominent features are the steepening (\emph{knee}) of the spectrum at about $4 \times 10^{15} \, \mathrm{eV}$ and its recovery to a harder slope (\emph{ankle}) at about $4 \times 10^{18} \, \mathrm{eV}$. The knee is due to a decrease in the flux of light primaries, as shown by the KASCADE experiment~\cite{Antoni2003490,Antoni20051} and others, e.g.~the EAS-TOP installation~\cite{aglietta2004,eastop04}.
In case of a rigidity dependent steepening in the spectra of the different primaries, the knee in the spectrum of the heavy component should appear at an energy of around $10^{17} \, \mathrm{eV}$~\cite{Hoerandel2004241,hillas06}. 
According to results from the KASCADE-Grande experiment~\cite{Apel2010202} a knee-like feature in the spectrum of heavy particles is visible 
at $E = 10^{16.92 \pm 0.04} \, \mathrm{eV}$, as recently published~\cite{PhysRevLett107}. 
It was shown that indeed the bending in the all-particle spectrum at $E = 10^{16.92 \pm 0.10} \, \mathrm{eV}$ is caused by a decrease in the flux of the heavy component. 
The combined spectrum of light and intermediate primaries, however, was therein found to be compatible with a single power law.

It is generally argued (see e.g.~\cite{Blasi12,Aloisio12}) that the transition from galactic to extragalactic origin of cosmic rays is expected in the energy range from $10^{17} \, \mathrm{eV}$ to $10^{19} \, \mathrm{eV}$~\cite{bergmann07}. In these models one should expect a hardening of the proton or, more generally, light primaries component of the cosmic ray spectrum to take place below or around $10^{18} \, \mathrm{eV}$ - due to the onset of the extragalactic contribution dominated by light primaries. The present study aims to search for experimental evidence for such a spectral feature. 
We use the same reconstruction and analysis procedures as described in~\cite{PhysRevLett107}, but with increased statistics and optimized selection criteria for enhancing light primaries.

\begin{figure}
\centering
\includegraphics[width=0.8\columnwidth]{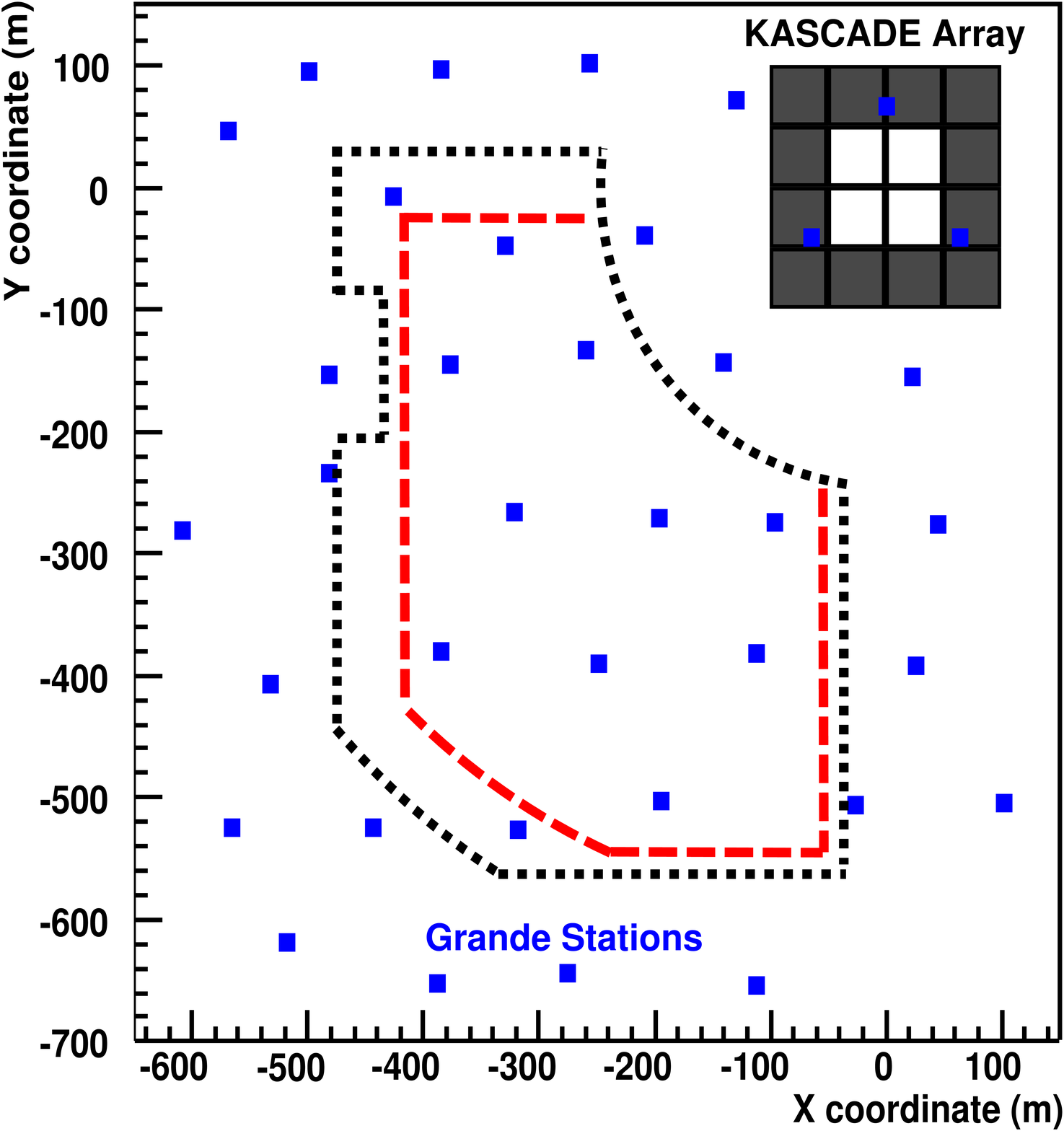}
\caption{(color online). Layout of the KASCADE-Grande experiment. The dashed line marks the fiducial area used in~\citep{PhysRevLett107}. The area used in this analysis is indicated by the dotted line. The experiment is located at KIT in Karlsruhe. The Grande stations (rectangles) had to be arranged to fit between the buildings. Therefore the stations are irregularly distributed and the missing station at around $(-600, \ -100)$ is the reason for the rectangular cut between $-80 \, \mathrm{m} \leq Y \leq -200 \, \mathrm{m}$. The original KASCADE array is located in the upper right corner. A comprehensive description of the detector and the reconstruction procedures can be found in~\cite{Apel2010202}.} 
\label{fig_grande_layout}
\end{figure}

The number of charged particles, $N_{\mathrm{ch}}$, and the number of muons, $N_{\mathrm{\upmu}}$, are the observables used in this analysis. Both values are defined as total numbers of particles reaching the observation level, where a lateral distribution function is fitted to the measured densities at the detector stations and integrated. The layout of the KASCADE-Grande experiment is shown in Fig.~\ref{fig_grande_layout}, where the Grande array records the number of charged particles exceeding an energy of $3 \, \mathrm{MeV}$ by means of 37 scintillation detector stations, distributed over an area of $700 \times 700 \, \mathrm{m^2}$. The total number of muons is determined by the measurements of shielded scintillation detectors at the KASCADE array, where the threshold energy for vertically incident muons is $230 \, \mathrm{MeV}$. A detailed description of the event reconstruction is given in~\cite{Apel2010202}.

A first step to increase statistics is to define a larger fiducial area than that used in the first analysis~\cite{PhysRevLett107} (see Fig.~\ref{fig_grande_layout}).
Since the probability to misreconstruct the number of muons increases with larger distances to the KASCADE detectors and for lower energies, events with core positions farther away from the KASCADE array are omitted resulting in the selected fiducial area. By increasing the fiducial area by $28 \, \mathrm{\%}$, we also shift the energy threshold to higher energies. This is no problem for the present analysis, because we are mainly interested in the development of the spectrum of light elements at energies well above threshold. After including additional 87 complete days of data taking, the total increase in statistics is about $ 36 \, \mathrm{\%}$ above energy of full efficiency. 

An adequate combination of $N_{\mathrm{ch}}$ and $N_{\mathrm{\upmu}}$ is used to estimate the primary energy. 
To classify the events into primary mass groups, we use the ratio of $N_{\mathrm{ch}}$ to $N_{\mathrm{\upmu}}$ in terms of the parameter \textit{k} (Eq.(\ref{form_k})). 
As a function of $\log_{10}N_{\mathrm{ch}}$, \textit{k} is centered around 0 for protons and it increases with the mass of the primary particle to become 1 for iron. In the same way, the reconstructed energy for a given $\log_{10}N_{\mathrm{ch}}$ increases with \textit{k} from the proton energy towards the energy for iron primaries.
The \textit{k} parameter is then used in the energy-calibration-function (Eq.(\ref{form_energy}), see also reference~\cite{Bertaina2012217}).
\begin{equation}
k = \frac{\log_{10}(N_{\mathrm{ch}}/N_{\mathrm{\upmu}})-\log_{10}(N_{\mathrm{ch}}/N_{\mathrm{\upmu}})_{\mathrm{H}}}{\log_{10}(N_{\mathrm{ch}}/N_{\mathrm{\upmu}})_{\mathrm{Fe}}-\log_{10}(N_{\mathrm{ch}}/N_{\mathrm{\upmu}})_{\mathrm{H}}},
\label{form_k}
\end{equation}
\begin{equation}
\log_{10}(N_{\mathrm{ch}}/N_{\mathrm{\upmu}})_{\mathrm{H,Fe}} = c_{\mathrm{H,Fe}} \cdot \log_{10}(N_{\mathrm{ch}}) + d_{\mathrm{H,Fe}},
\end{equation} 
\begin{multline}
\log_{10}(E/\mathrm{GeV}) = (a_{\mathrm{H}} + (a_{\mathrm{Fe}} - a_{\mathrm{H}}) \cdot k) \cdot \log_{10}(N_{\mathrm{ch}})\\ + b_{\mathrm{H}} + (b_{\mathrm{Fe}} - b_{\mathrm{H}}) \cdot k
\label{form_energy}
\end{multline} 
$a_{\mathrm{H,Fe}}$, $b_{\mathrm{H,Fe}}$, $c_{\mathrm{H,Fe}}$ and $d_{\mathrm{H,Fe}}$ are obtained by fitting linear functions to the mean $\log_{10}E_{\mathrm{true}}$ (coefficients $a$ and $b$) and to the mean $\log_{10}(N_{\mathrm{ch}} / N_{\mathrm{\upmu}})$ (coefficients $c$ and $d$) of simulated events as a function of the logarithm of their reconstructed number of charged particles. This has been done separately for 5 zenith angle intervals of equal exposure with the upper limits $16.7^\circ$, $24^\circ$, $29.9^\circ$, $35.1^\circ$ and $40^\circ$, to take the shower attenuation into account. The parameters are slightly different from those used in~\cite{PhysRevLett107}, because they are determined using the larger fiducial area and taking into account the higher energy threshold. For the simulations of the air showers, the CORSIKA code~\cite{CORSIKA} was used employing the QGSJET-II-2 high energy hadronic interaction model~\cite{qgsjet06}. Fluka (version 2002.4)~\cite{fluka} was used for interactions at low energies. The systematic/statistical uncertainties in determining the all-particle spectrum as well as the energy resolution of the present analysis are shown in Table~\ref{table_systematics}.

\begin{table}[htbp]
\caption{Energy resolution and systematic and statistical uncertainties for the all-particle spectrum are given for three different energies. A more detailed discussion of the uncertainties can be found in reference~\cite{Apel2012183}.}
\begin{tabular}{|l|r|r|r|}
\hline
 & \multicolumn{1}{l|}{$10^{16.6} \, \mathrm{eV}$} & \multicolumn{1}{l|}{$10^{17.0} \, \mathrm{eV}$} & \multicolumn{1}{l|}{$10^{17.8} \, \mathrm{eV}$} \\ \hline
syst. uncertainty & 11.6$ \, \%$ & 11.1$ \, \%$ & 24.4$ \, \%$ \\ \hline
statistical error & 0.5$ \, \%$ & 1.2$ \, \%$ & 12.0$ \, \%$ \\ \hline
energy resolution & 22.1$ \, \%$ & 20.5$ \, \%$ & 16.1$ \, \%$ \\ \hline
\end{tabular}
\label{table_systematics}
\end{table}

\begin{figure}
\centering
\includegraphics[width=0.8\columnwidth]{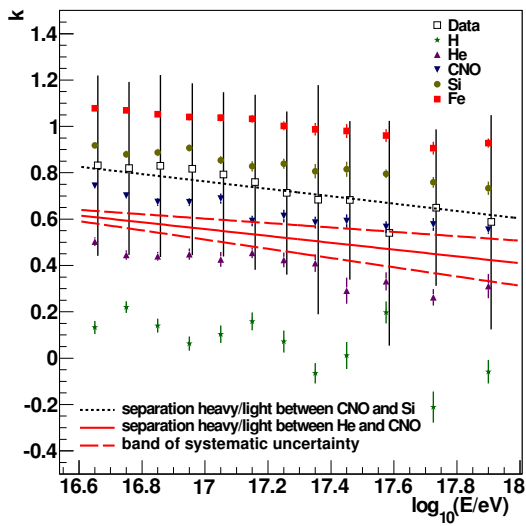}
\caption{(color online). The mean values of \textit{k} over the reconstructed energy are shown for events simulated with QGSJet-II-2, and with zenith angles between $0^{\circ}-24^{\circ}$, and five different primaries. 
The distributions of \textit{k} over \textit{E} are similar in adjacent zenith angle intervals. By combining two angular bins, we increase simulation statistics. 
The increase of systematic uncertainties is taken into account. 
For comparison, the measured data (empty squares) is also shown (shifted from the bin center to the right for better visibility of the error bars). The error bars represent the RMS for the measured data and the error of the mean for the simulated data. 
The dotted line was used in~\cite{PhysRevLett107} to separate the events into a light and a heavy component. 
For an enhanced light selection on cosmic rays, the continuous line is taken for the separation, where the dashed lines depict the uncertainty of the separation, taking into account also the reconstruction uncertainty of \textit{k}.}
\label{fig_kcut}
\end{figure}

In the following, \textit{k} is used to separate the all-particle energy spectrum into two mass groups: the heavy (electron-poor) and the light (electron-rich) component. This is done by fitting a separation line to the simulated mean \textit{k}-values over the reconstructed energy and comparing the calculated \textit{k}-value of each event with the value given by the separation line. The distributions of \textit{k} for 5 different masses are shown in Fig.~\ref{fig_kcut}. Shower-to-shower fluctuations are larger for lighter primaries. The dotted line was used in~\cite{PhysRevLett107} to separate the events into mass groups. The line is obtained by fitting the $k_{\mathrm{sep}}(E)=[k_{\mathrm{CNO}}(E)/2 + k_{\mathrm{Si}}(E)/2]$ distribution, i.e.~between the CNO and Si simulations. 
In the present analysis a lower \textit{k}-cut is used to enhance the light component, which consists of events with assigned \textit{k}-values smaller than the values shown as solid line (obtained by a fit to $k_{\mathrm{sep}}(E)=[k_{\mathrm{He}}(E)/2 + k_{\mathrm{CNO}}(E)/2]$). 
The dashed lines, finally, represent the estimate of a possible error of the separation, taking also the reconstruction uncertainty of \textit{k} into account. 
They are obtained by shifting $k_{\mathrm{sep}}(E)$ up/down by the statistical and systematic uncertainties of \textit{k} before the fit is performed. A more detailed description of the energy determination procedures and the separation into mass groups can be found in \cite{PhysRevLett107} and \cite{Apel2012183}.  

Mainly due to the shower-to-shower fluctuations, it is not possible to derive spectra for individual elements using \textit{k} on an event-by-event basis. However, it should be noted that in the relevant energy range these fluctuations are nearly energy independent. By separating the events into light and heavy mass groups according to the present procedure, we can assume that these components contain at least most of the light/heavy individual elements, respectively, according to the hadronic interaction model employed.
\begin{figure}
\centering
\includegraphics[width=\columnwidth]{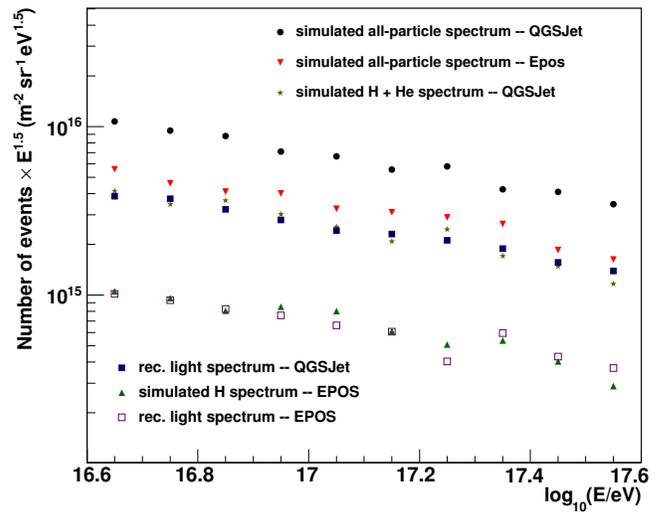}
\caption{(color online). Shown are simulated and reconstructed all-particle energy spectra using QGSJet-II and EPOS-1.99 as hadronic interaction models, where for EPOS a smaller number of events are available. 
For the simulations, a composition of five elements (H, He, CNO, Si, and Fe) equally abundant has been used. 
The combined proton and helium component for QGSJet-II and the pure hydrogen component for EPOS are also displayed. 
In addition, reconstructed spectra for the light component (separation between He and CNO) are shown, where in both cases the QGSJet based reconstruction and selection criteria are applied. In case of QGSJet-II the light component is reproduced, whereas for EPOS simulations the reconstructed light component shows an agreement with the initial pure proton spectrum.} \label{fig_epos_qgs}
\end{figure}

A simple attempt to cross-check the model dependence is displayed in Fig.~\ref{fig_epos_qgs}, where we reconstruct events generated with EPOS (version 1.99 \cite{epos09}) applying the QGSJet calibration. The spectra have been simulated with a slope index of -2 and weighted to $E^{-3}$ to better represent the index of the measured data in this energy range. For the simulations, a composition of five elements (H, He, CNO, Si, and Fe) with equal abundances has been used. The reconstructed light spectra show a significant difference in composition, where EPOS generated data result in a much lighter composition. This is probably caused by the fact that EPOS predicts more muons compared to QGSJet-II and, therefore, the ratio of $N_{\mathrm{ch}}$ to $N_{\mathrm{\upmu}}$ is smaller for a given number of charged particles resulting in a larger \textit{k}-value. Especially helium events migrate (by calibrating with QGSJet-II) to the heavy mass group. This effect might be slightly compensated by the higher reconstructed energy of the events~\cite{mario2012}. Using an EPOS calibration, the measured showers appear to originate from lighter primaries and of lower energy compared to the QGSJet-II calibration.
Figure~\ref{fig_epos_qgs} also demonstrates that the selection of events according to the $k$-parameter does not induce any artificial structures in the spectra of light primaries. If the data are well described by QGSJet-II, then the spectrum of light primaries with the separation between He and CNO should consist mainly of protons and helium, maybe with some additional, less abundant elements between helium and carbon. This can be seen in Fig.~\ref{fig_epos_qgs}, where the combined simulated proton and helium component for QGSJet-II is in good agreement with the reconstructed spectrum of light elements, which has been obtained by applying the QGSJet-II based reconstruction and selection criteria to the data simulated using QGSJet-II. Assuming that the data simulated with EPOS are closer to real data, then the measured spectrum of light particles is an almost pure proton spectrum.
The simulated proton spectrum for EPOS is similar to the reconstructed spectrum of light primaries, which has been derived from EPOS generated events using again the QGSJet-II based reconstruction and selection criteria. According to QGSJet-II, the spectrum of heavy elements for the same separation would contain carbon and primaries heavier than that. For EPOS it should also contain most of the helium component.

\begin{figure}
\centering
\includegraphics[width=\columnwidth]{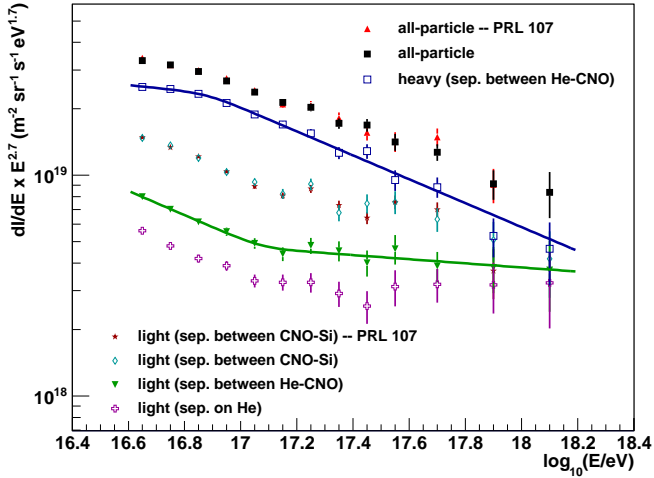}
\caption{(color online). The all-particle and electron-rich spectra from the analysis~\cite{PhysRevLett107} in comparison to the results of this analysis with higher statistics. In addition to the light and heavy spectrum based on the separation between He and CNO, the light spectrum based on the separation on He is also shown. The error bars show the statistical uncertainties.} \label{fig_vgl_prl}
\end{figure}
\begin{figure}
\centering
\includegraphics[width=\columnwidth]{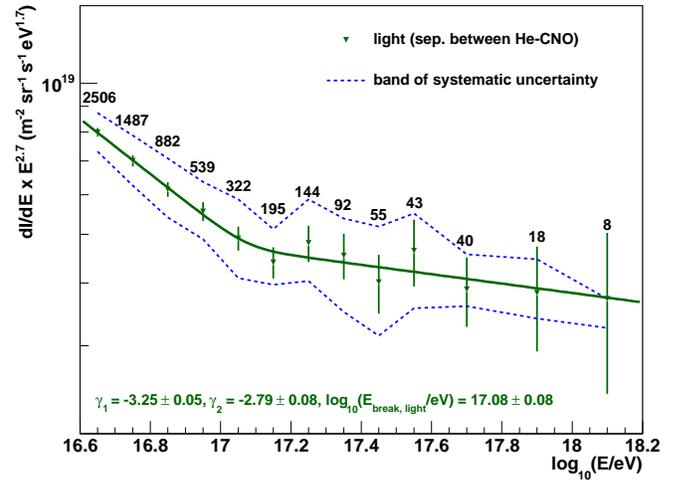}
\caption{(color online). The reconstructed energy spectrum of the light mass component of cosmic rays.  
The number of events per energy bin is indicated as well as the range of systematic uncertainty. The error bars show the statistical uncertainties.} \label{fig_light}
\end{figure}

In Fig.~\ref{fig_vgl_prl}, the results of the present analysis are shown. To cross-check the results from~\cite{PhysRevLett107} the all-particle spectrum and the spectrum of light primaries for the former used area and data are compared with the ones obtained with higher statistics from the present studies. Both all-particle spectra and spectra of light elements based on the separation between CNO and Si are in good agreement. The spectra of light and heavy particles with the separation between He and CNO are obtained using the separation-line shown in Fig.~\ref{fig_kcut}. 
The spectrum of the heavy component, which now contains also the medium mass component, exhibits a change of index at $E = 10^{16.88 \pm 0.03} \, \mathrm{eV}$ and it therefore agrees inside the corresponding uncertainty with the previous result~\cite{PhysRevLett107} at $E^{\mathrm{heavy}}_{\mathrm{knee}} = 10^{16.92 \pm 0.04} \, \mathrm{eV}$. The hardening or ankle-like feature visible in the enriched spectrum of light primaries is more prominent compared to the one that includes the CNO component. Although statistics gets quite low for the spectrum of light elements with the separation on He (obtained by a fit to the mean \textit{k}-values for He in Fig.~\ref{fig_kcut}), it is obvious that it cannot be described by one single power law only. Formula~(\ref{form_fitfunc})~\cite{TerAntonyan:2000hh} is used for fitting the spectra of the light and heavy components:
\begin{equation}
\begin{split}
&\frac{dI}{dE}(E) = I_{0} \cdot E^{\gamma_{1}} \cdot [1 + (\frac{E}{E_{\mathrm{b}}})^{\epsilon}]^{(\gamma_{1} - \gamma_{2})/\epsilon},\\
&\text{$I_{0}$ : normalization factor},\\
&\text{$\gamma_{1/2}$ : index before/after the bending},\\
&\text{$E_{\mathrm{b}}$ : energy of the break position},\\
&\text{$\epsilon$ : smoothness of the break.}\\
\end{split}
\label{form_fitfunc}
\end{equation}

As shown in Fig.~\ref{fig_light}, a change of the spectral index from $\gamma_{1} = -3.25 \pm 0.05$ to 
$\gamma_{2} = -2.79 \pm 0.08$ at an energy of $10^{17.08 \pm 0.08} \, \mathrm{eV}$ is observed for the light component. The dashed lines mark the systematic error band for the separation between He and CNO obtained by using the selection shown in Fig.~\ref{fig_kcut}.
The measured number of events above the bending is $N_{\mathrm{meas}} = 595$. Without the bending we would expect $N_{\mathrm{exp}} = 467$ events above this ankle-like feature. The Poisson probability to measure at least $N_{\mathrm{meas}}$ events above the bending, if $N_{\mathrm{exp}}$ events are expected, is $P(N\geq N_{\mathrm{meas}}) = \sum_{k = N_{\mathrm{meas}}}^{\infty}(\frac{N_{\mathrm{exp}}^{k}}{k!} e^{(-N_{\mathrm{exp}})}) \approx 7.23 \times 10^{-09}$. This corresponds to a significance of $5.8 \, \mathrm{\sigma}$ that in this energy range the spectrum of light primaries cannot be described by a single power law.
If we shift the separation criteria in order to obtain an even purer proton sample (sep. on He, fig.~\ref{fig_vgl_prl}) the slope difference even increases ($\gamma_{1} = -3.32 \pm 0.03$ to $\gamma_{2} = -2.59 \pm 0.28$ at an energy of $10^{17.16 \pm 0.19} \, \mathrm{eV}$).

So far, possible uncertainties due to the underlying hadronic interaction model have not been taken into account. But, as discussed in~\cite{donghwa2012}, applying different methods in energy calibration does not result in different conclusions on changes of the spectral slopes. Spectral slopes do not significantly change, when corrections for the bin-to-bin fluctuations (i.e. by unfolding methods correcting the reconstruction uncertainties) are applied~\cite{Apel2012183}. Therefore, it is not expected that such a correction would  result in a spectrum of light primaries compatible with a single power law.

One important observation is that the knee in the heavy component occurs at a lower energy compared to the bending in the spectrum of light primaries. This is still the case, if the break positions are shifted by their uncertainties. Therefore, the steepening of the heavy spectrum and the recovery of the light component is not due to a bias in the reconstruction or separation procedures. In addition, such a bias should result in a bending in the spectrum of light elements at a lower energy compared to the spectrum of heavy particles: a bias in the separation would be visible in the spectrum of light particles first, because its flux is much smaller. A bias in the reconstruction, where the calculated \textit{k}-value is too low, would result in a reconstruction of heavy particles as light particles with a lower energy.
It is worth to point out that the slope of the heavy mass spectrum above the knee-like feature is very similar to the slope of the light mass spectrum before the ankle-like feature. This could be a hint to a similar source population, as described e.g.~in~\cite{Kalmykov1997}. Correspondingly, the slope index of $\gamma_2 = -2.7$ might be an indication of an injection spectrum of a new (extragalactic) source population of high energy cosmic rays~\cite{Aloisio12,Blasi12,bergmann07}.

In summary, after redefining what is considered as a electron-poor/electron-rich event and increasing the statistics, a significant hardening, i.e.~an ankle-like feature in the cosmic ray spectrum of light primaries, is observed at 
$E = 10^{17.08 \pm 0.08} \, \mathrm{eV}$. The slope index of the underlying power law changes at this energy from $-3.25 \pm 0.05$ to 
$-2.79 \pm 0.08$, which might be an indication that the transition from galactic to extragalactic origin of cosmic rays starts already in this energy range.

\begin{acknowledgments}
The authors would like to thank the members of the engineering and technical staff of the KASCADE-Grande collaboration, who contributes to the success of the experiment. The KASCADE-Grande experiment was supported by the BMBF of Germany, the MIUR and INAF of Italy, the Polish Ministry of Science and Higher Education, and the Romanian Authority for Scientific Research
UEFISCDI (PNII-IDEI grants 271/2011 and 17/2011). J.C.A.V acknowledges the partial support of CONACyT and the DAAD-Proalmex program (2009-2012).
The present study is supported by the 'Helmholtz Alliance for Astroparticle Physics - HAP' funded by the Initiative and Networking Fund of the Helmholtz Association, Germany.
\end{acknowledgments}


%

\end{document}